\documentclass[twoside,english]{elsarticle}
\usepackage[T1]{fontenc}
\usepackage{mathrsfs}
\usepackage{amsmath}
\usepackage{amssymb}
\usepackage{graphicx}

\makeatletter
\journal{editor}

\makeatother

\usepackage{babel}
\begin{document}

\begin{frontmatter}{}

\title{Unitary evolution to a state with a fixed mean number of particles}

\author{Bogus\l aw Broda}

\ead{boguslaw.broda@uni.lodz.pl}

\ead{http://merlin.phys.uni.lodz.pl/BBroda}

\address{Department of Theoretical Physics, Faculty of Physics and Applied
Informatics, University of \L \'{o}d\'{z}, 90-236 \L \'{o}d\'{z},
Pomorska 149/153, Poland}
\begin{abstract}
In the framework of finite-dimensional Fock space models, for a predefined
fixed mean number of particles $\bar{n}_{k}$, it is shown that there
is a ``large'' multi-dimensional subspace $s_{\bar{n}_{k}}$ of
initial pure states, in the space $S$ of all pure states, unitarily
evolving to a subspace $S_{\bar{n}_{k}}$ of final pure states which
yield $\bar{n}_{k}$. As an example, in particular it follows that
the blackbody form of the mean number of particles $\bar{n}_{k}$
does not by itself contradict unitarity of black hole evaporation.
\end{abstract}
\begin{keyword}
states with a fixed mean number of particles \sep unitary evolution
in finite-dimensional Hilbert spaces \sep unitarity of black hole
evaporation

\PACS 03.65.Aa \texttt{(Quantum systems with finite Hilbert space)}
\sep 04.70.Dy \texttt{(Quantum aspects of black holes, evaporation,
thermodynamics)} \MSC[2010] 81Q99 \texttt{(None of the above, but
in this section)} \sep 83C57 \texttt{(Black holes)}
\end{keyword}

\end{frontmatter}{}

\section{Introduction}

The aim of the work is to show that there are many (initial) pure
states unitarily evolving to a (pure) state with a predefined fixed
mean number of particles $\bar{n}_{k}$. Although primary motivation
to pose such a problem has resulted from analysis of the black hole
information paradox \citep{Chakraborty2017,Harlow2016a,Marolf2017a,Polchinski2017c},
actually the problem directly refers to quantum mechanics and is entirely
independent from black hole context.

In short, the black hole information (loss) paradox (problem/puzzle)
consists in difficulty in explaining the status of unitarity of the
process of black hole evaporation. According to standard picture of
semiclassical gravity, a black hole (quantumly) evaporates due to
the Hawking effect, and finally transmutes into blackbody radiation.
In consequence, in general distinct initial pure states forming a
black hole could possibly be transformed into the same final ``structureless''
blackbody (thermal) radiation. Due to this seemingly ``many to one''
process, unitarity (of evolution) could be lost in contradistinction
with fundamentals of quantum mechanics. 

In the present paper we analyze the issue of unitarity of the process
of evolution of a pure state to a (pure) state with a predefined fixed
mean number of particles $\bar{n}_{k}$. We present our arguments
in three steps, in the form of the following three models defined
in finite-dimensional Hilbert spaces: (1) Toy Model (a model defined
on the Bloch sphere \citep{Bengtsson2017,Chruscinski2004,Nielsen2000});
(2) a more realistic, Fermion Fock space model; and (3) Boson Fock
space model with a cutoff \citep{Trzetrzelewski2004a}. More precisely,
we will show that there is a ``large'' (in the sense of dimension)
subspace $S_{\bar{n}_{k}}$ of distinct pure states in the whole space
$S$ of pure states (or in the corresponding Hilbert space $\mathcal{H}$)
yielding the same (almost arbitrary) \emph{mean number of particles}
$\bar{n}_{k}$. Therefore, a process which could seem, at first glance,
to be ``many to one'' can actually be ``one to one'' (and unitary),
because there is ``enough room'' in the space $S$ to ``accommodate''
this process.

One should stress that we exclusively operate on pure states, i.e.,
nowhere do density matrices, nor mixed states or purifications of
thereof appear in our considerations, explicitly or implicitly. In
our analysis, $\bar{n}_{k}$, blackbody or another, is a mean of the
particle number operator $\hat{n}_{k}$ in a pure state (see (\ref{eq:n-as-oper.average-between-nstates})),
and possible thermality of the spectrum is, in a sense, simulated
by an appropriately chosen pure state $\left|\left.\bar{n}_{k}\right\rangle \right.$
(see (\ref{eq:final-nstate-on-Bloch}), (\ref{eq:tensorial-Bloch}),
(\ref{eq:boson-tensorial-Bloch})).

In the context of black holes, we do not prove that the actual process
of black hole evaporation is ``one to one'' and unitary, but as
an illustration of our analysis we argue that the blackbody(-like)
shape of the Hawking spectrum does not by itself imply non-unitarity
of black hole evaporation. Obviously, our analysis is quite general,
because no explicit particular form of the mean number of particles
$\bar{n}_{k}$ (blackbody or another) enters our analysis.

\section{Primary motivation and the Toy Model}

In his famous work on black hole radiation spectrum \citep{Hawking1975a},
Hawking derived an explicit formula for the mean number of particles
$\bar{n}_{k}$ understood as a quantum average
\begin{equation}
\bar{n}_{k}\equiv\left\langle \hat{n}_{k}\right\rangle ,\label{eq:def.of-no.n-as-average}
\end{equation}
where $\hat{n}_{k}$ is the particle number operator for the mode
$k$. With a good approximation, the mean number of particles of black
hole radiation, $\bar{n}_{k}$, appears to be blackbody. Actually,
what counts from our perspective is the \emph{total mean number of
particles} \citep{Broda2017}, rather than usually discussed temporary
quantity, but as it is mentioned in Introduction our analysis is insensitive
to any particular form of $\bar{n}_{k}$.

\subsection{General idea}

Our main claim is that we have a ``huge multitude'' of pure states
yielding a predefined fixed mean number of particles $\bar{n}_{k}$.
More precisely, we have a ``large'', in the sense of low codimension,
subspace denoted by $S_{\bar{n}_{k}}$ in the space $S$ of all pure
states (or in the Hilbert space $\mathcal{H}$) corresponding to almost
any arbitrarily chosen $\bar{n}_{k}$ (the only restriction on $\bar{n}_{k}$
is imposed, depending on the case, by the mild condition (\ref{eq:small-interval}),
(\ref{eq:general-small-interval}) or (\ref{eq:boson-small-interval})).
Therefore, having given the mean $\bar{n}_{k}$, and provided we are
able to determine the corresponding (``large'') subspace $S_{\bar{n}_{k}}$,
we choose any state $\left|\bar{n}_{k}\right\rangle \in S_{\bar{n}_{k}}$
yielding, by virtue of the definition of $S_{\bar{n}_{k}}$, the average
with the expected predefined value(s), i.e.,
\begin{equation}
\left\langle \bar{n}_{k}\left|\hat{n}_{k}\right|\bar{n}_{k}\right\rangle =\bar{n}_{k}.\label{eq:n-as-oper.average-between-nstates}
\end{equation}
In the next step we can perform a (thought) unitary transformation
$U\left(-t\right)$ on $S_{\bar{n}_{k}}$, now interpreted as a subspace
of final states, obtaining another subspace $s_{\bar{n}_{k}}$ (incidentally,
because of unitarity of $U\left(-t\right)$, $s_{\bar{n}_{k}}$ is
isometric to $S_{\bar{n}_{k}}$ in the sense of the complex metric
on $\mathcal{H}$) interpreted as a subspace of possible initial states.
The unitary transformation $U\left(-t\right)$ corresponds to evolution
backward in time (the minus sign). Thus, we can conclude that the
``huge multitude'' of distinct initial pure states belonging to
$s_{\bar{n}_{k}}$ unitarily (according to $U\left(t\right)$) evolves
towards the ``huge multitude'' of distinct pure states belonging
to the subspace $S_{\bar{n}_{k}}$ of states yielding, by virtue of
the construction, the fixed predefined mean number of particles $\bar{n}_{k}$.

\subsection{Toy Model}

Now, let us introduce the Toy Model (a model on the Bloch sphere).
Its sole role is to explicitly elucidate and visualize (because of
low dimension) our main idea. As a chief postulate of the model we
assume that the whole Universe consists of only one fermion mode (2-level
system). Its Hilbert space $\mathcal{H}=\mathbb{C}^{2}$ is 4-dimensional
in real sense (in this paper we only operate real dimensions), and
in the Fock space base $\left\{ \left|0\right\rangle ,\left|1\right\rangle \right\} $
any state $\left|\psi\right\rangle \in\mathcal{H}$ can be expressed
as
\begin{equation}
\left|\psi\right\rangle =\alpha_{0}\left|0\right\rangle +\alpha_{1}\left|1\right\rangle ,\qquad\alpha_{0},\alpha_{1}\in\mathbb{C}.\label{eq:general-state-in-C2}
\end{equation}
Because of normalization ($\left\langle \psi|\psi\right\rangle =1$)
and of arbitrariness of phase, pure states for this system are parameterized
by points on the 2-dimensional Bloch sphere $\mathcal{S}^{2}\left(=\mathbb{C}P^{1}\right)$
\citep{Bengtsson2017,Chruscinski2004,Nielsen2000}. Then, the general
state can be specified as
\begin{equation}
\left|\psi\right\rangle =\cos\frac{\theta}{2}\left|0\right\rangle +e^{i\varphi}\sin\frac{\theta}{2}\left|1\right\rangle ,\qquad0\leq\theta\leq\pi,\quad0\leq\varphi<2\pi,\label{eq:general-state-on-Bloch}
\end{equation}
where $\theta$ and $\varphi$ are the polar and azimuthal angle on
$\mathcal{S}^{2}$, respectively. From the point of view of quantum
mechanics, any (pure) state is uniquely given by a point on $\mathcal{S}^{2}$,
and arbitrary continuous unitary (e.g.\ time) evolution $U\left(t\right)$
corresponds to rotation of $\mathscr{\mathcal{S}}^{2}$, i.e., $U\left(t\right)\in SO\left(3\right)$.

Now, we would like to determine the entire (sub)space $S_{\bar{n}}$
of the states $\left|\bar{n}\right\rangle $ yielding the fixed mean
number of particles $\bar{n}$, where obviously
\begin{equation}
0\leq\bar{n}\leq1.\label{eq:interval}
\end{equation}
Since $k=1$, the mode number $k$ has been skipped in this subsection,
and the mean number of particles $\bar{n}$ is now a single number
belonging to the interval (\ref{eq:interval}). In general, the state
we are looking for, expressed as
\begin{equation}
\left|\bar{n}\right\rangle =\bar{\alpha}_{0}\left|0\right\rangle +\bar{\alpha}_{1}\left|1\right\rangle \label{eq:nstate-in-C2}
\end{equation}
(in our paper the ``bar'' symbol over coefficients denotes their
specific values, instead, complex conjugation is denoted by the ``asterisk''),
should satisfy the two conditions:
\begin{equation}
\left\langle \bar{n}|\bar{n}\right\rangle =1,\qquad\left\langle \bar{n}\right|\hat{n}\left|\bar{n}\right\rangle =\bar{n}.\label{eq:def.nstate}
\end{equation}
Utilizing the angular parametrization (\ref{eq:general-state-on-Bloch})
on the Bloch sphere, we easily find
\begin{equation}
\bar{n}=\left|\bar{\alpha}_{1}\right|^{2}=\sin^{2}\frac{\theta}{2}.\label{eq:cond.on-abar1}
\end{equation}
Thus, finally the solution of the problem (full set of solutions of
Eqs.(\ref{eq:def.nstate}) modulo phase) assumes the following explicit
form
\begin{equation}
\left|\bar{n}\right\rangle =\sqrt{1-\bar{n}}\left|0\right\rangle +e^{i\varphi}\sqrt{\bar{n}}\left|1\right\rangle .\label{eq:final-nstate-on-Bloch}
\end{equation}
Eq.(\ref{eq:final-nstate-on-Bloch}) says that all the points (interpreted
by us as final pure states) on the circle (``parallel'') $S_{\bar{n}}$,
parameterized by the azimuthal angle $\varphi$ and determined by
the ``latitude'' (on $\mathcal{S}^{2}$) given by the polar angle
(see (\ref{eq:cond.on-abar1})) $\theta=2\arcsin\sqrt{\bar{n}}$,
yield the same predefined $\bar{n}$. In the next step we can (thought)
rotate the circle (parallel) $S_{\bar{n}}$ (``backward in time evolution
$U\left(-t\right)$'') obtaining another circle (not a parallel,
in general) $s_{\bar{n}}$ parametrizing all the states (interpreted
as initial states) which can be transformed back (in the course of
the ``proper time evolution $U\left(t\right)$'') onto states on
$S_{\bar{n}}$ with the predefined $\bar{n}$. Thus, in general, we
have a ``one to one'' unitary relation between points on isometric
circles on $\mathcal{S}^{2}$, where parallels play a distinguished
role of ``thermality imitating'' states. Since for $\bar{n}=0,1$
the circle $S_{\bar{n}}$ degenerates to a point (poles), we can impose
a mild restriction on admissible values of $\bar{n}$, removing the
boundary values of the interval (\ref{eq:interval}), putting
\begin{equation}
0<\bar{n}<1.\label{eq:small-interval}
\end{equation}

\begin{figure}[h]
\begin{centering}
\includegraphics[scale=0.35]{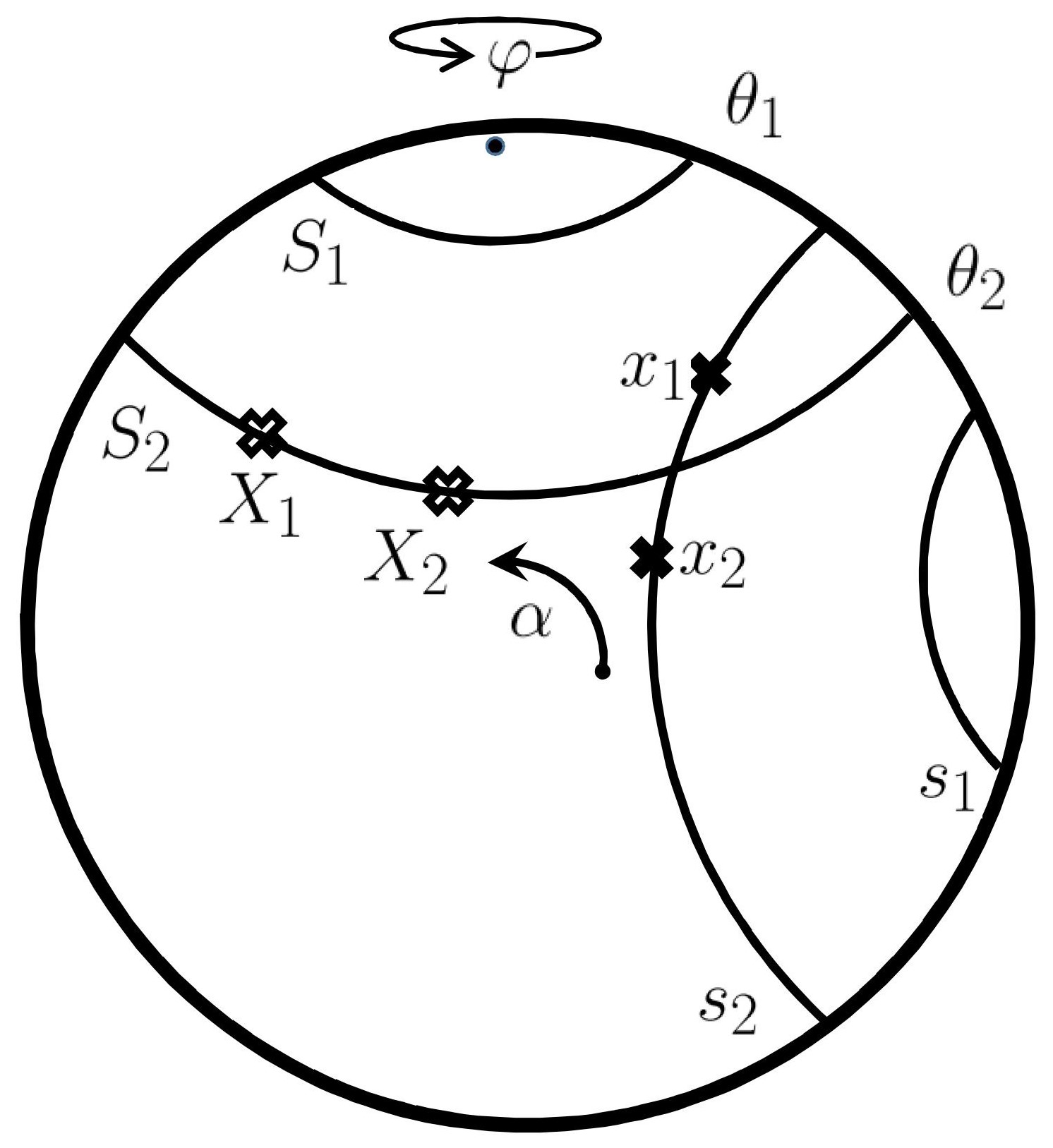}
\par\end{centering}
\caption{Toy Model (a model on the Bloch sphere $\mathcal{S}^{2}$). \emph{Geometric
statement}: Rotation $U\left(\alpha\right)$ through the angle $\alpha=\pi/2$
in the plane of Figure transforms the circles $s_{1}$ and $s_{2}$
onto the circles $S_{1}$ and $S_{2}$, respectively. Polar angle
coordinates of the circles $S_{1}$ and $S_{2}$ are $\theta_{1}$
and $\theta_{2}$, respectively, and the points on the circles are
parametrized by the azimuthal angle $\varphi$. In particular, the
distinct points $x_{1},x_{2}\in s_{2}$ are transformed (rotated)
onto the points $X_{1},X_{2}\in S_{2}$, respectively.\emph{ Quantum-mechanical
statement}: Two distinct initial pure states $x_{1}$ and $x_{2}$
unitarily (after ``time $\pi/2$'') evolve onto two distinct pure
states $X_{1},X_{2}$, respectively, with the same mean number of
particles (``spectrum'') $\bar{n}=\sin^{2}\frac{\theta_{2}}{2}$.}
\end{figure}

An example situation is explicitly illustrated in Fig.1.

\section{Fock space models}

In this section we introduce two more realistic models based on fermion
Fock space and boson Fock space, respectively. To technically simplify
our discussion (algebraization of the problem), as well as to make
it more quantitative and rigorous, we impose some cutoffs on the Fock
spaces, which implies finite dimensionality of corresponding Hilbert
spaces.

\subsection{Fermion Fock space model}

First, we consider a fermion model defined on the antisymmetric Fock
space $\mathcal{F}_{A}^{m}$, where $m$~($\geq1$) is a number of
fermion modes. Here, the cutoff simply means that the number of modes
$m$ is finite. Generalizing the linear expansion (\ref{eq:general-state-in-C2}),
we can express any state $\left|\psi\right\rangle \in\mathcal{F}_{A}^{m}$
as a linear combination
\begin{equation}
\left|\psi\right\rangle =\mathop{\sum_{\boldsymbol{n}}\alpha_{\boldsymbol{n}}\left|\boldsymbol{n}\right\rangle },\qquad\alpha_{\boldsymbol{n}}\in\mathbb{C},\label{eq:general-psi}
\end{equation}
where for convenience we have introduced a multi-index $\boldsymbol{n}\equiv n_{1}\cdots n_{m}$
with $n_{k}=0,1$ ($k=1,\ldots,m$).

Analogously to the presentation (\ref{eq:nstate-in-C2}), the points/states
$\left|\bar{\boldsymbol{n}}\right\rangle \in\tilde{S}_{\bar{\boldsymbol{n}}}$,
i.e., those normalized and satisfying the condition (\ref{eq:n-as-oper.average-between-nstates}),
can be expressed by the sum
\begin{equation}
\left|\bar{\boldsymbol{n}}\right\rangle =\mathop{\sum_{\boldsymbol{n}}}\bar{\alpha}_{\boldsymbol{n}}\left|\boldsymbol{n}\right\rangle ,\label{eq:general-nbar}
\end{equation}
where for the multi-parameter $\bar{\boldsymbol{n}}\equiv\bar{n}_{1},\ldots,\bar{n}_{m}$
we assume $0\leq\bar{n}_{k}\leq1$, and the ``tilde'' over $S_{\bar{\boldsymbol{n}}}$
denotes the space of states before identification (symbolized by ``$/\sim$'')
of states differing by phase, i.e., $S_{\bar{\boldsymbol{n}}}\equiv\tilde{S}_{\bar{\boldsymbol{n}}}/\sim$.
Henceforth, we use the symbol $S_{\bar{\boldsymbol{n}}}$ instead
of $S_{\bar{n}_{k}}$. Normalization condition for (\ref{eq:general-nbar})
reads
\begin{equation}
\left\langle \bar{\boldsymbol{n}}|\bar{\boldsymbol{n}}\right\rangle =\mathop{\sum_{\boldsymbol{n}}}\left|\bar{\alpha}_{\boldsymbol{n}}\right|^{2}=1,\label{eq:general-normalization}
\end{equation}
whereas the condition (\ref{eq:n-as-oper.average-between-nstates})
yields the following system of $m$ quadratic equations
\begin{equation}
\left\langle \bar{\boldsymbol{n}}\left|\hat{n}_{k}\right|\bar{\boldsymbol{n}}\right\rangle =\mathop{\sum_{\boldsymbol{n}_{k}}}\left|\bar{\alpha}_{\boldsymbol{n}_{k}}\right|^{2}=\bar{n}_{k},\label{eq:general-average}
\end{equation}
where we have introduced another multi-index $\boldsymbol{n}_{k}\equiv n_{1}\cdots1_{k}\cdots n_{m}$,
i.e., the $k$th index assumes the constant value $n_{k}=1$, and
consequently there is no summation with respect to $n_{k}$ in (\ref{eq:general-average}).
Furthermore, a bit extending the domain of the index $k$ introducing
a new auxiliary index $p=0,1,\ldots,m$ instead of $k$ ($=1,\ldots,m$),
and additionally defining $\boldsymbol{n}_{0}\equiv\boldsymbol{n}$
as well as $\bar{n}_{0}\equiv1$, we can rewrite the quadratic equation
(\ref{eq:general-normalization}) and the system (\ref{eq:general-average})
in the following compact unified form:
\begin{equation}
\mathop{\sum_{\boldsymbol{n}_{p}}}\left|\bar{\alpha}_{\boldsymbol{n}_{p}}\right|^{2}=\bar{n}_{p},\qquad p=0,1,\ldots m.\label{eq:compact-eqations}
\end{equation}

Thus, the (sub)space $\tilde{S}_{\bar{\boldsymbol{n}}}$ is defined
as a (sub)space of solutions of the system of $m+1$ quadratic equations
(\ref{eq:compact-eqations}). Fortunately, to proceed further we do
not need an explicit form of $S_{\bar{\boldsymbol{n}}}$, as in the
case of the Toy Model of Section 2, where global analysis has been
performed. Since we only aim to determine the (co)dimension of $\tilde{S}_{\bar{\boldsymbol{n}}}$
(and of $S_{\bar{\boldsymbol{n}}}$), we can confine ourselves to
purely local analysis.

Our strategy is first to find only a single non-degenerate (in the
sense explained latter) solution of the quadratic system (\ref{eq:compact-eqations}),
and next to show that it can be infinitesimally extended in sufficiently
many dimensions/directions. It is straightforward to check that the
following ``(tensor product) Bloch-type'' state (its symmetrized
version is known as the Dicke state \citep{Dicke1954,Martin2010})
\begin{equation}
\left|\bar{\boldsymbol{n}};\boldsymbol{\phi}\right\rangle \equiv\mathop{\sum_{\boldsymbol{n}}\bar{\alpha}_{\boldsymbol{n}}}\left(\boldsymbol{\phi}\right)\left|\boldsymbol{n}\right\rangle ,\qquad\boldsymbol{\phi}=\varphi_{1},\ldots,\varphi_{m},\quad0\leq\varphi_{k}<2\pi\quad\left(k=1,\ldots,m\right),\label{eq:tensorial-Bloch}
\end{equation}
where (cf.\,(\ref{eq:general-state-on-Bloch}))
\begin{equation}
\bar{\alpha}_{\boldsymbol{n}}\left(\boldsymbol{\phi}\right)\equiv\bar{\alpha}_{n_{1}\cdots n_{m}}\left(\varphi_{1},\ldots,\varphi_{m}\right)=\mathop{\prod_{k=1}^{m}}\left(\delta_{n_{k}}^{0}\cos\frac{\theta_{k}}{2}+\delta_{n_{k}}^{1}e^{i\varphi_{k}}\sin\frac{\theta_{k}}{2}\right),\label{eq:a-bar-phi}
\end{equation}
with (cf.\,(\ref{eq:cond.on-abar1}))
\begin{equation}
\sin^{2}\frac{\theta_{k}}{2}=\bar{n}_{k},\label{eq:sin2-theta}
\end{equation}
solves the system (\ref{eq:compact-eqations}). Actually, the formulas
(\ref{eq:tensorial-Bloch}\textendash \ref{eq:sin2-theta}) define
the whole $m$-dimensional torus $\mathcal{T}^{m}\equiv\underset{m}{\underbrace{\mathcal{S}^{1}\times\cdots\times\mathcal{S}^{1}}}$
of solutions of the system (\ref{eq:compact-eqations}), parameterized
by $\boldsymbol{\phi}$. Since our analysis is supposed to be local,
we only need a single point/solution of the system (\ref{eq:compact-eqations}),
and therefore, to simplify our further considerations we put $\boldsymbol{\phi}=0$
henceforth.

To find a solution of the system (\ref{eq:compact-eqations}) in an
infinitesimal vicinity of the Bloch-type solution (\ref{eq:tensorial-Bloch}\textendash \ref{eq:sin2-theta})
at the point $\boldsymbol{\phi}=0$ on the torus $\mathcal{T}^{m}$,
we insert (into (\ref{eq:compact-eqations})) the expansion
\begin{equation}
\bar{\alpha}_{\boldsymbol{n}}=\bar{\alpha}_{\boldsymbol{n}}\left(0\right)+z_{\boldsymbol{n}},\label{eq:a-plus-z}
\end{equation}
where $z_{\boldsymbol{n}}$ is a (complex) infinitesimal variation
around the solution $\bar{\alpha}_{\boldsymbol{n}}\left(0\right)$.
Thus, we get a system of $m+1$ linear equations
\begin{equation}
\textrm{Re}\mathop{\sum_{\boldsymbol{n}_{p}}}\bar{\alpha}_{\boldsymbol{n}_{p}}\left(0\right)z_{\boldsymbol{n}_{p}}^{*}=0,\qquad p=0,1,\ldots,m,\label{eq:Re-linear}
\end{equation}
which define $m+1$ hyperplanes tangent to the respective $m+1$ quadrics
determined by the system (\ref{eq:compact-eqations}). The maximal
possible rank of the matrix of the coefficients entering the system
(\ref{eq:Re-linear}) is obviously $m+1$, and such a situation (the
most desirable one) geometrically corresponds to a non-degenerate
intersection of the hyperplanes tangent to the quadrics. Since to
determine the rank of a matrix, one usually invokes determinants,
let us calculate the determinant of a matrix constructed from the
columns containing the following coefficients: $\bar{\alpha}_{00\cdots0}\left(0\right)$,
$\bar{\alpha}_{10\cdots0}\left(0\right)$, $\bar{\alpha}_{01\cdots0}\left(0\right)$,
..., $\bar{\alpha}_{00\cdots1}\left(0\right)$. The matrix reads
\begin{equation}
M_{A}=\left[\begin{array}{cccc}
\bar{\alpha}_{00\cdots0}\left(0\right) & * & \cdots & *\\
 & \bar{\alpha}_{10\cdots0}\left(0\right)\\
 &  & \ddots\\
 &  &  & \bar{\alpha}_{00\cdots1}\left(0\right)
\end{array}\right],\label{eq:matrix-M-A}
\end{equation}
where $M_{A}$ appears to be an upper triangular matrix (more precisely,
all entries of $M_{A}$, possibly except the 1st row and the main
diagonal, are zero). Then, by virtue of (\ref{eq:a-bar-phi})

\begin{equation}
\det M_{A}=\bar{\alpha}_{00\cdots0}\left(0\right)\mathop{\prod_{j=1}^{m}}\bar{\alpha}_{00\cdots1_{j}\cdots0}\left(0\right)=\mathop{\prod_{j=1}^{m}}\sin\frac{\theta_{j}}{2}\cos^{m}\frac{\theta_{j}}{2}.\label{eq:det}
\end{equation}
From (\ref{eq:det}) it immediately follows that the rank of the system
(\ref{eq:Re-linear}) is really maximal ($=m+1$), and hence there
is no degeneracy, provided we impose the condition $0<\theta_{k}<\pi$,
which corresponds (by virtue of the relationship (\ref{eq:sin2-theta}))
to a very mild restriction on $\bar{n}_{k}$ (cf.\,(\ref{eq:small-interval})),
\begin{equation}
0<\bar{n}_{k}<1,\qquad k=1,\ldots,m,\label{eq:general-small-interval}
\end{equation}
in comparison with all theoretically admissible values: $0\leq\bar{n}_{k}\leq1$.

\subsection{Boson Fock space model}

Let us now switch to a boson model defined on the symmetric Fock space
$\mathcal{F}_{S}^{m,N}$ with cutoffs $m$ ($\geq1$) and $N$ ($\geq1)$,
where $m$ is a finite number of boson modes, and a fixed finite number
of possible levels, the same for each boson mode, is equal to $N+1$
(see \citep{Trzetrzelewski2004a}). In principle the cutoff $N$ can
be arbitrary, but for our needs it should be sufficiently large, i.e.,
\begin{equation}
\bar{n}_{k}<N<+\infty,\qquad k=1,\ldots,m.\label{eq:N-inequality}
\end{equation}

As a single-mode Bloch-type state (with $\varphi=0$) in the cutoff
boson case ($\mathcal{F}_{S}^{m,N}$) we assume
\begin{equation}
\left|\psi\right\rangle =\cos\frac{\theta}{2}\left|0\right\rangle +\sin\frac{\theta}{2}\left|N\right\rangle ,\label{eq:boson-Bloch-type}
\end{equation}
where $0\leq\theta\leq\pi$ (cf.\ (\ref{eq:general-state-on-Bloch})).
Executing calculations similar to those in Section 2, we obtain as
analog of (\ref{eq:cond.on-abar1})
\begin{equation}
\bar{n}=N\sin^{2}\frac{\theta}{2}.\label{eq:boson-theta}
\end{equation}

In turn, the tensor product Bloch-type state (with $\boldsymbol{\phi}=0$)
is now (cf.\,(\ref{eq:tensorial-Bloch}))
\begin{equation}
\left|\bar{\boldsymbol{n}};0\right\rangle =\mathop{\sum_{\boldsymbol{n}}\bar{\alpha}_{\boldsymbol{n}}\left(0\right)\left|\boldsymbol{n}\right\rangle },\label{eq:boson-tensorial-Bloch}
\end{equation}
where (cf.\,(\ref{eq:a-bar-phi}))
\begin{equation}
\bar{\alpha}_{\boldsymbol{n}}\left(0\right)=\mathop{\prod_{k=1}^{m}\left(\delta_{n_{k}}^{0}\cos\frac{\theta_{k}}{2}+\delta_{n_{k}}^{N}\sin\frac{\theta_{k}}{2}\right)},\label{eq:boson-abar}
\end{equation}
with (cf.\,(\ref{eq:sin2-theta}) and (\ref{eq:general-boson-theta}))
\begin{equation}
\sin^{2}\frac{\theta_{k}}{2}=\frac{\bar{n}_{k}}{N}.\label{eq:general-boson-theta}
\end{equation}
In the boson case, for the multi-index $\boldsymbol{n}$ we assume
$n_{k}=0,1,\ldots N$, and for the multi-parameter $\boldsymbol{\bar{n}}$,
$0\leq\bar{n}_{k}<+\infty$, respectively.

The matrix analogous to (\ref{eq:matrix-M-A}) is now the (upper triangular)
matrix
\begin{equation}
M_{S}=\left[\begin{array}{cccc}
\bar{\alpha}_{00\cdots0}\left(0\right) & * & \cdots & *\\
 & \bar{\alpha}_{N0\cdots0}\left(0\right)\\
 &  & \ddots\\
 &  &  & \bar{\alpha}_{00\cdots N}\left(0\right)
\end{array}\right],\label{eq:matrix-M-S}
\end{equation}
and its determinant,

\begin{equation}
\det M_{S}=\mathop{\prod_{j=1}^{m}}\sin\frac{\theta_{j}}{2}\cos^{m}\frac{\theta_{j}}{2},\label{eq:boson-determinant}
\end{equation}
is exactly of the same form as for $M_{A}$ (see (\ref{eq:det})).
The relations (\ref{eq:boson-theta}) and (\ref{eq:boson-determinant})
impose a very mild restriction on $\bar{n}_{k}$ (cf.\,(\ref{eq:general-small-interval})),
\begin{equation}
0<\bar{n}_{k}<+\infty,\qquad k=1,\ldots,m,\label{eq:boson-small-interval}
\end{equation}
in comparison with all theoretically admissible values: $0\leq\bar{n}_{k}<+\infty$.

\subsection{Summary of the Fock space models}

In the case of the fermion Fock space model as well as in the case
of the boson one, for arbitrary fixed mean number of particles $\bar{n}_{k}$,
mildly restricted by (\ref{eq:general-small-interval}) and (\ref{eq:boson-small-interval}),
respectively, we have shown that the tensor product Bloch-type state
(\ref{eq:tensorial-Bloch}\textendash \ref{eq:sin2-theta}) and (\ref{eq:boson-tensorial-Bloch}\textendash \ref{eq:general-boson-theta}),
respectively, determines a non-degenerate intersection point in the
corresponding (finite dimensional) Hilbert space $\mathcal{H=F}_{A}^{m}$
and $\mathcal{H=F}_{S}^{m,N}$, respectively. More precisely, intersecting
hyperplanes tangent to the intersecting quadrics defined by the system
(\ref{eq:compact-eqations}) (with indices $n_{k}=0,1$ and $n_{k}=0,1,\ldots,N$,
respectively) are ``in general position'' (genuine intersection,
no overlappings). Therefore, the intersection of the quadrics is also
non-degenerate (genuine intersection, no contact points) and hence
each quadric (equation in the system of the $m+1$ equations (\ref{eq:compact-eqations}))
imposes one condition reducing dimension of the subspace by one (codimension
increases by one). Consequently, codimension of the intersection of
the whole set of the quadrics (\ref{eq:compact-eqations}) equals
codimension of the intersection of the set of the hyperplanes tangent
to these quadrics, and it is equal $m+1$ (number of the equations).
Then, $\dim\tilde{S}_{\bar{\boldsymbol{n}}}=\dim\mathcal{H}-(m+1)$
and 
\begin{equation}
\dim S_{\bar{\boldsymbol{n}}}=\dim\tilde{S}_{\bar{\boldsymbol{n}}}-1=\dim\mathcal{H}-m-2\label{eq:dimsn}
\end{equation}
$\left(S_{\bar{\boldsymbol{n}}}\equiv\tilde{S}_{\bar{\boldsymbol{n}}}/\sim\right)$.
Strictly speaking, the subtraction in (\ref{eq:dimsn}) is justified
provided the action of the group $U\left(1\right)$ corresponding
to the identification (``$\sim$'') of states differing by phase
proceeds tangentially to $\tilde{S}_{\bar{\boldsymbol{n}}}$ at the
point $\bar{\alpha}_{\boldsymbol{n}}\left(0\right)$. Glancing at
the system (\ref{eq:Re-linear}) we can immediately observe that imaginary
parts ($y_{\boldsymbol{n}}\equiv\textrm{Im }z_{\boldsymbol{n}}$)
of infinitesimal variations $z_{\boldsymbol{n}}$ are absent from
the system (\ref{eq:Re-linear}) (unrestricted), and therefore (infinitesimally)
$\tilde{S}_{\bar{\boldsymbol{n}}}$ can freely extend in imaginary
directions, and this is exactly the direction of (infinitesimal) action
of the (phase) group $U\left(1\right)$.

For illustrative purposes, let us apply Eq.(\ref{eq:dimsn}) to the
simplest non-trivial case, namely, to our Toy Model (Subsection 2.2).
Since now $\dim\mathcal{H}=4$ and $m=1$, we get $\dim S_{\bar{n}}=4-1-2=1,$
which obviously agrees with dimension of a circle (``parallel'').

To better quantify and justify the term ``huge multitude'' or ``large''
introduced in Section 2 in the context of dimension of $S_{\bar{\boldsymbol{n}}}$,
we should compare dimensions of all relevant spaces. To begin with,
for the Hilbert space of states $\mathcal{H}$ we have $\dim\mathcal{H}=2\left(N+1\right)^{m}$,
where $N=1$ or $N=\textrm{cutoff}$ (see (\ref{eq:N-inequality})),
in the fermion case ($\mathcal{F}_{A}^{m}$) or in the boson case
($\mathcal{F}_{S}^{m,N}$), respectively. Since the normalization
condition for states lowers dimension by one, and so does also identification
of states differing by phase \citep{Bengtsson2017,Chruscinski2004},
(dimension of the space $S$ of all states) $\dim S=\dim\mathcal{H}-2=2\left(N+1\right)^{m}-2$.
In turn, according to (\ref{eq:dimsn}) $\dim S_{\bar{\boldsymbol{n}}}=2\left(N+1\right)^{m}-m-2$.
Then, $S_{\bar{\boldsymbol{n}}}$ is a {[}$2\left(N+1\right)^{m}-m-2${]}-dimensional
subspace in the {[}$2\left(N+1\right)^{m}-2${]}-dimensional space
$S$ of all states, and consequently codimension of $S_{\bar{\boldsymbol{n}}}$
in $S$ is equal to the number $m$ of the modes. Then, $\textrm{codim}\thinspace S_{\bar{\boldsymbol{n}}}=m\sim C\log\dim S$
(where $C=\log^{-1}\left(N+1\right)$) and exactly in this (``logarithmic'')
sense is dimension of $S_{\bar{\boldsymbol{n}}}$ large.

Interpreting $S_{\bar{\boldsymbol{n}}}$ as a subspace of final (pure)
states with a fixed spectrum, and performing a (thought) unitary transformation
$U\left(-t\right)$ (understood as evolution backward in time), we
obtain (an isometric to $S_{\bar{\boldsymbol{n}}}$ subspace) $s_{\bar{\boldsymbol{n}}}$,
which can be interpreted as a subspace of initial states. By virtue
of the construction, unitary (time) evolution determined by $U\left(t\right)$
transforms all pure states belonging to $s_{\bar{\boldsymbol{n}}}$
onto (pure) states belonging to $S_{\bar{\boldsymbol{n}}}$ which
yield the same predefined fixed mean number of particles $\bar{n}_{k}$.

\section{Final remarks}

In the presented analysis we have shown that there is a huge multitude
of distinct pure states which can unitarily be evolved to states with
a predefined fixed mean number of particles $\bar{n}_{k}$. Comparing
in the framework of our finite-dimensional Fock space models a subspace
$S_{\bar{\boldsymbol{n}}}$ of final pure states yielding $\bar{n}_{k}$,
e.g.\ blackbody-like or another, to the whole space $S$ of pure
states, we observe that asymptotically the order of growth of codimension
of $S_{\bar{\boldsymbol{n}}}$ ($\subset S$) is (actually) only logarithmic
function of dimension of $S$. Then, as $S_{\bar{\boldsymbol{n}}}$
is a (sub)space which parametrizes a really ``huge multitude'' of
all final states with given $\bar{n}_{k}$, and $s_{\bar{\boldsymbol{n}}}$
(isometric to $S_{\bar{\boldsymbol{n}}}$) is a (sub)space of all
possible initial states, there is ``enough room'' in the space $S$
to ``accommodate'' unitarily realized time evolution.

To illustrate this result in the context of black hole evaporation,
we should adopt the following scenario. In the beginning of evolution,
as an input state we have matter forming a black hole in a pure state.
In turn, the final state only consists of the radiation (in a pure
state) which is solely characterized by its blackbody(-like) spectrum
(mean number of particles) $\bar{n}_{k}$. Since according to our
analysis the space of all possible input pure states yielding the
given $\bar{n}_{k}$ is large, the black hole evaporation can be unitarily
realized with an ease. Actually, black hole radiation is characterized
not only by its blackbody spectrum but also by thermal density matrix.
Therefore, for preserving unitarity, not only the blackbody spectrum
$\bar{n}_{k}$ should be reproduced properly (and in principle it
can), but so should also the averages of all observables. Obviously,
this later property cannot be reproduced by any pure state.

I am grateful for Reviewer's very accurate and valuable remarks.

\bibliographystyle{elsarticle-num}
\bibliography{unitary_evolution_to_a_state_with_a_predefined_mean_number_v08}

\end{document}